\documentclass[12pt,preprint]{aastex}

\usepackage{natbib}
\usepackage{epsfig}

\newcommand{\rhill}{r_{\mathrm{H}}}
\newcommand{\msun}{\mathrm{M}_{\sun}}
\newcommand{\mearth}{\mathrm{M}_{\oplus}}
\newcommand{\AU}{\:\mathrm{AU}}
\newcommand{\figref}[1]{Figure \ref{#1}}
\newcommand{\eqref}[1]{equation (\ref{#1})}

\shorttitle{Radiative Transfer in Protoplanetary Disks}
\shortauthors{Jang-Condell \& Sasselov}
\slugcomment{submitted to ApJ February 18, 2003 -- revised \today}

\begin{document}
\newcommand{\etal}{et al.~}
\newcommand{\eg}{e.g.~}
\title{Radiative Transfer on Perturbations in Protoplanetary Disks}
\author{Hannah Jang-Condell\altaffilmark{1}
\and Dimitar D.~Sasselov\altaffilmark{2}}
\affil{Harvard-Smithsonian Center for Astrophysics}
\affil{60 Garden St., Cambridge, MA 02138}
\altaffiltext{1}{hjang@cfa.harvard.edu}
\altaffiltext{2}{dsasselov@cfa.harvard.edu}

\newcommand{\tunpert}{133.6\:\mathrm{K}}	
% 1:       130.683      135.475      133.596      135.475
\newcommand{\tsmmin}{130.7\:\mathrm{K}}
\newcommand{\tsmmax}{135.5\:\mathrm{K}}
\newcommand{\tsmd}{5\:\mathrm{K}}
\newcommand{\losm}{2.2}
\newcommand{\hism}{1.4}
\newcommand{\dsm}{4}
% 10:      123.678      142.308      133.596      157.056
\newcommand{\tlgmin}{123.7\:\mathrm{K}}
\newcommand{\tlgmax}{142.3\:\mathrm{K}}
\newcommand{\tlgd}{19\:\mathrm{K}}
\newcommand{\lolg}{7.4}
\newcommand{\hilg}{6.5}
\newcommand{\dlg}{14}
% 20:      122.036      146.574      133.596      153.426
\newcommand{\txlmin}{122.0\:\mathrm{K}}
\newcommand{\txlmax}{146.6\:\mathrm{K}}
\newcommand{\txld}{25\:\mathrm{K}}
\newcommand{\hixl}{8.7}
\newcommand{\loxl}{9.7}
\newcommand{\dxl}{18}

\begin{abstract}
We present a method for calculating the radiative 
tranfer on a protoplanetary disk perturbed by a protoplanet.  
We apply this method to determine the effect on the temperature 
structure within the photosphere of a passive circumstellar disk 
in the vicinity of a small protoplanet of up to 20 Earth masses.  
The gravitational potential of a protoplanet induces a compression 
of the disk material near it, resulting in a decrement in the density 
at the disk's surface.  Thus, an isodensity contour at the height 
of the photosphere takes on the shape of a well.  When such a well is 
illuminated by stellar irradiation at grazing incidence, 
it results in cooling in a shadowed 
region and heating in an exposed region.  For typical stellar and 
disk parameters relevant to the epoch of planet formation, 
we find that the temperature variation due to a protoplanet at 
1 AU separation from its parent star is about \dsm\% ($\tsmd$) for a planet 
of 1 Earth mass, about \dlg\% ($\tlgd$) for planet of 10 Earth masses, 
and about \dxl\% ($\txld$) for planet of 20 Earth masses, 
We conclude that even such relatively small protoplanets can induce 
temperature variations in a passive disk.  Therefore, many of the processes 
involved in planet formation should not be modeled with a locally 
isothermal equation of state.
\end{abstract}


\section{Introduction}

Planetary systems are formed from rotating protoplanetary disks,
which are often modeled
in steady-state, in vertical hydrostatic equilibrium, with gas and 
dust fully mixed and thermally coupled \citep{kenyon}.  
Such disks describe very well the observed properties of
T Tauri disks of age $\sim$1~Myr and typical mass accretion rates
of $\leq 10^{-8} M_{\odot} {\rm yr}^{-1}$ \citep{hartmann,vertstruct}. 
They are passive disks in the sense modeled
by \eg \citet{CG}, where the main source of photospheric heating 
is irradiation from the parent star, although viscous heating is 
still important at small stellar distances and near the midplane.  
The temperature in such disks
is computed under the assumption that the upper surface of the disk
is perfectly concave and smooth at all radii, which is a very good
description of an unperturbed disk because thermal and
gravitational instabilities are damped very efficiently \citep{thermstab}.

We are interested in planet formation inside such disks, which
means that we might be forced to abandon the above assumption of
a perfectly smooth surface. The newly formed planet core will
distort it, affect the heating and cooling of the disk
locally, and could have significant consequences for the further
growth and migration of protoplanets. As discussed in \citet{sas_lec}, 
the distortion need only be large enough compared to
the grazing angle at which the starlight strikes the disk. This
small angle has a minimum
at $0.4\AU$ and increases significantly only at very large distances.
The depth of the depression due to the additional mass of the planet, $m_p$,
will be proportional to $(\rhill/h)^3$, where $\rhill$ is the Hill radius,
and $h$ is the local scale height, with
a shade area dependent on the grazing angle \citep{sas_lec}.

Generally planet formation has been treated numerically in 2-dimensional
disks; to date, there is only one recent study of the 3-dimensional
effect due to planets embedded in a protoplanetary disk \citep{bate}.
Bate \etal model the 3-dimensional response of a gaseous viscous disk
with a simple locally isothermal equation of state, without taking 
into account irradiation from the central star.  

In this paper we study in detail the development of a compression
in a standard passive disk and the radiative transfer in it. 
We consider this to be the first step in developing a global
3-dimensional simulation, similar to the one by \citet{bate},
but with a realistic equation of state, especially near and inside
the protoplanet's Roche lobe.
We are explicitly interested in the planetary growth process before a 
gap in the disk is formed, hence our approach is limited to 
$m_p\lesssim30\:\mearth$, which results in a small perturbation to the 
irradiated surfaces of a passive disk.

\section{Heating Sources}

\begin{figure}
\plotone{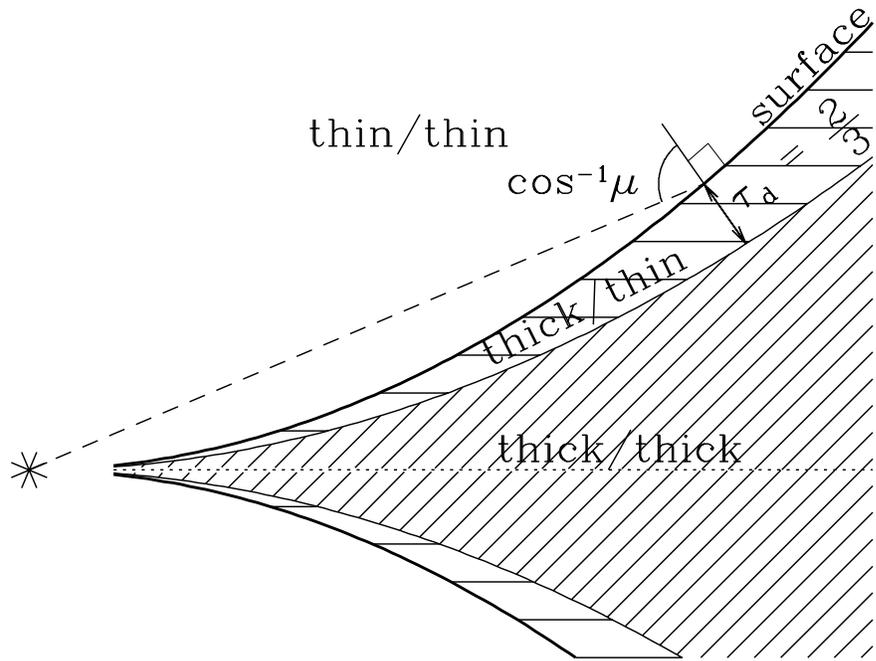}
\caption{\label{flareddisk}Schematic of a flared disk. 
Proportions are not to scale.  The labels thick and thin 
refer to optical depth in short and long wavelengths.
}
\end{figure}

A schematic of our adopted disk model is shown in \figref{flareddisk}.  
The disk is flared, so that disk height increases faster than 
distance to the star, consistent with 
observational evidence and theoretical models
\citep{kenyon,CG,vertstruct}.  
We assume that the dust is well mixed with the gas 
and is the primary source of 
opacity at wavelengths characteristic of stellar emission.  
The dust absorbs incident stellar radiation at the surface 
and re-emits at longer wavelength, where the opacity is much lower.  
We are interested in the region of the disk that is optically thick, 
at distances of $\lesssim10$ AU \citep{dalessio2}.  
This gives rise to three different layers with smooth transitions 
in the disk: 
the uppermost is optically thin to both stellar and disk radiation, 
the interior is optically thick to both stellar and disk radiation, 
and the middle layer is a transition region that is optically thick 
to stellar radiation, but still optically thin to disk radiation.  
We shall refer to this optically thick/thin region as the photosphere.  
The reprocessing of stellar radiation in this region 
is important for determining the interior temperature.  

\citet{CG} calculate the temperature of only the thin/thin and
thick/thick regions using energy conservation considerations, without
detailing the temperature structure of the photosphere.
\citeauthor{vertstruct} (\citeyear{vertstruct},\citeyear{dalessio2})
calculate the detailed temperature structure of the
photosphere in disks around young stars in a self-consistent way,
including radiative transport and convection. 
They find that radiative transfer is the dominant mechanism for energy 
transport in the disk photosphere, and the primary heating sources are 
stellar irradiation and viscous
heating, as shown by comparison to \citet{calvet}, who use the
Milne-Strittematter treatment of the superposition of solutions
\citep{milne,strittmatter}.  We shall consider stellar radiation and
viscous heating to be the primary sources of heating, where the total
temperature is arrived at by calculating the sum of the viscous and
radiative fluxes, as \(T^4 = T_v^4 + T_r^4\), where $T_v$ is
the viscous temperature and $T_r$ is stellar radiation heating
temperature.  

\subsection{Viscous Heating}

The temperature due to viscous heating for a constant flux, gray atmosphere is 
\begin{equation}
T_v^4 = \frac{3F_v}{4\sigma_B}(\tau_d+2/3)
\end{equation}
where $\sigma_B$ is the Stefan-Boltzmann constant, 
$\tau_d$ is the optical depth of the disk's own radiation,
and the viscous flux $F_v$ at a distance $a$, for a star of effective 
temperature $T_{\star}$, mass $M_{\star}$, and radius $R_{\star}$, 
accreting at a rate $\dot{M}_a$ is 
\begin{equation}
F_v = \frac{3GM_{\star}\dot{M}_a}{4\pi a^3}
	\left[1-\left(\frac{R_{\star}}{a}\right)^{1/2}\right]
\end{equation}
\citep{pringle}.

\subsection{Stellar Irradiation}

We shall use a modified form of the Milne-Strittmatter
treatment to calculate the effect of heating by stellar irradiation 
on a three-dimensional
perturbation on a passive accretion disk.
We assume that the dust is well-mixed with the gas in the disk and that 
they are thermodynamically coupled so that the dust temperature 
and gas temperature are the same.  In general, there is no guarantee
that these two temperatures are the same, or that the dust is either 
well mixed or uniform in composition.

Following the Milne-Strittmatter treatment, the incident radiation 
from the star is considered to be at some characteristic wavelength, 
typically in the visible, and the radiation emitted by the 
disk will be at a longer characteristic wavelength, typically 
in the infrared.  
The incident, short wavelength radiation will be indicated by the 
subscript $s$ (stellar or scattered), and the re-emitted, long wavelength 
radiation will be indicated by the subscript $d$ (disk or diffuse).
Some fraction of the energy of the incident stellar 
radiation, $\sigma$, is scattered 
at the same frequency.  The remaining fraction, $\alpha = 1-\sigma$, is 
absorbed by dust in the disk and re-emitted at a longer characteristic 
wavelength, typically in the infrared.  This gives the 
equation of radiative transfer for the 
stellar short wavelength radiation,
\begin{equation}
\frac{1}{\kappa_s \rho} \mathbf{\hat{k}} \cdot \nabla I_s =
	I_s - \frac{\sigma E_0 e^{-\tau_s}}{4\pi}
	- \frac{\sigma \int I_s d\Omega}{4\pi} 
\end{equation}
where $\rho$ is the density, $\kappa_s$ is the opacity at short wavelengths, 
and $E_0$ is the incident energy flux given by 
\begin{equation}
E_0 = \sigma_B T_{\star}^4 \left(\frac{R_{\star}}{a}\right)^2.
\end{equation}
The zeroth moment of this equation is 
\begin{equation}\label{s0}
\frac{1}{\kappa_s \rho} \nabla \cdot \mathbf{F}_s 
= \alpha 4\pi J_s - \sigma E_0 e^{-\tau_s},
\end{equation}
and the first moment, making use of the Eddington approximation, is
\begin{equation}\label{s1}
\frac{1}{\kappa_s \rho} \nabla J_s = \frac{3\,\mathbf{F}_s}{4\pi}. 
\end{equation}

The radiative transfer equation for the radiation absorbed 
and re-emitted within the disk is 
\begin{equation}
\frac{1}{\kappa_d\rho} \mathbf{\hat{k}} \cdot \nabla I_d = I_d - B
\end{equation}
where $\kappa_d$ is the optical depth at disk radiation wavelengths.  
The zeroth and first moments (using the Eddington approximation) are 
\begin{equation}\label{d0}
\frac{1}{\kappa_d\rho} \nabla \cdot \mathbf{F}_d = 4\pi (J_d - B),
\end{equation}
\begin{equation}\label{d1}
\frac{1}{\kappa_d\rho}\,\nabla J_d = \frac{3\,\mathbf{F}_d}{4\pi}.
\end{equation}

\subsubsection{Plane-parallel Disk}
\label{plane-parallel}

In a plane parallel disk atmosphere, this problem is reduced 
to one dimension, and we can express everything as a function of 
optical depth normal to the surface, $\tau_d$.  We can substitute 
\[ \frac{1}{\kappa_d\rho}\nabla \rightarrow \frac{d}{d\tau_d}
\quad \mbox{and} \quad
\mathbf{F}_{s,d} \rightarrow F_{s,d}. \]
If the angle of incidence 
of the stellar radiation is $\cos^{-1}\mu_0$, then 
\begin{equation}
\tau_d = \tau_s\mu_0/q
\end{equation}
where $q$ is the ratio of opacities
\begin{equation}
q \equiv \frac{\kappa_s}{\kappa_d}.
\end{equation}
The condition of zero net flux relates the short and long wavelength 
fluxes as 
\begin{equation}\label{sd}
F_s + F_d = E_0 \mu_0 \exp(-q\tau_d/\mu_0)
\end{equation}
and the boundary conditions at the surface are that of an isotropic 
radiation field, 
\begin{equation}
2\pi\,J_s(0) = F_s(0) \quad \mbox{and} \quad
2\pi\,J_d(0) = F_d(0).
\end{equation}

Now, equations (\ref{s0}), (\ref{s1}), (\ref{d0}), (\ref{d1}), and 
(\ref{sd}) are a closed system of linear differential equations 
that we can solve directly for the temperature, 
$T_r = (\pi B/\sigma_B)^{1/4}$.  The solution is given in 
\citet{calvet} as 
\begin{equation}
\label{radtemp}
B = \frac{\alpha E_0 \mu_0}{4\pi} 
[C_1' + C_2' \exp(-q\tau_d/\mu_0) + C_3' \exp(-\beta q\tau_d)]
\end{equation}
where $\beta\equiv\sqrt{3\alpha}$, 
\begin{eqnarray}
C_1' &=& (1+C_1)\left(2+\frac{3\mu_0}{q}\right) 
	+ C_2\left(2+\frac{3}{\beta q}\right), \\
C_2' &=& \frac{(1+C_1)}{\mu_0} \left(q-\frac{3\mu_0^2}{q}\right), \\
C_3' &=& C_2 \beta \left(q - \frac{3}{q\beta^2}\right),
\end{eqnarray}
and 
\begin{eqnarray}
C_1 &=& -\frac{3\sigma\mu_0^2}{1-\beta^2\mu_0^2}, \\
C_2 &=& \frac{\sigma(2+3\mu_0)}{\beta(1+2\beta/3)(1-\beta^2\mu_0^2)}.
\end{eqnarray}

\begin{figure}
\plotone{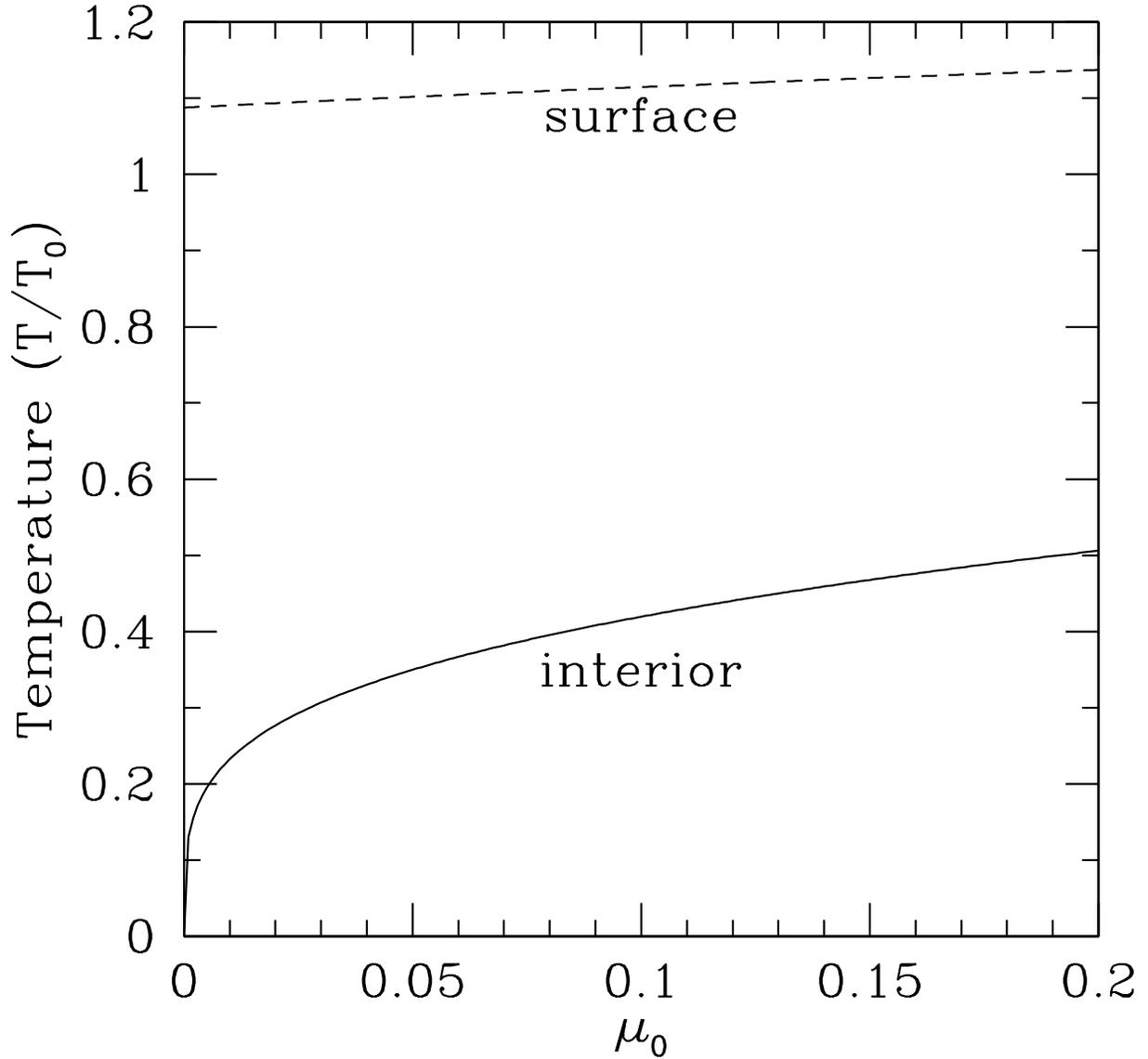}
\caption{\label{t_unperturbed}The temperature due solely to stellar 
irradiation at the surface of the disk (dashed line) 
and in the interior (solid line) 
for varying values of the angle of
incidence, $\mu_0$, where $T_0 = (E_0/\sigma_B)^{1/4}$.  
}
\end{figure}

\figref{t_unperturbed} shows how temperature varies with 
$\mu_0$ at the surface ($\tau_d=0$) and in the interior 
($\tau_d\rightarrow\infty$).  The temperatures are normalized 
to an effective radiant temperature $T_0 = (E_0/\sigma_B)^{1/4}$, 
which is the blackbody temperature corresponding to the incident energy flux.  
Note that at the surface the temperature remains relatively constant 
with $\mu_0$ and is 
greater than $T_0$ due to the absorptive properties of the dust, 
in accordance with the ``superheated'' surface layer proposed in \citet{CG}.
In contrast, the interior temperature is sensitive to the incident angle, 
especially at grazing incidence.  Thus, perturbations in the surface 
that change the angle of incidence can significantly affect 
the interior temperature.

By definition, $\tau_d=2/3$ is where the disk becomes optically 
thick to its own radiation.  
For $q\gtrsim1$, the temperature at $\tau_d=2/3$ is 
can be approximated by the interior temperature, since as 
$q\tau_d$ becomes large, the exponentials in \eqref{radtemp}
vanish and $B$ becomes independent of $\tau_d$.  
This indicates that in the absence of viscous heating, 
the disk interior is vertically isothermal for $\tau_d\geq2/3$.

\subsubsection{Disk with Perturbation}

\begin{figure}
\epsscale{0.5}
\plotone{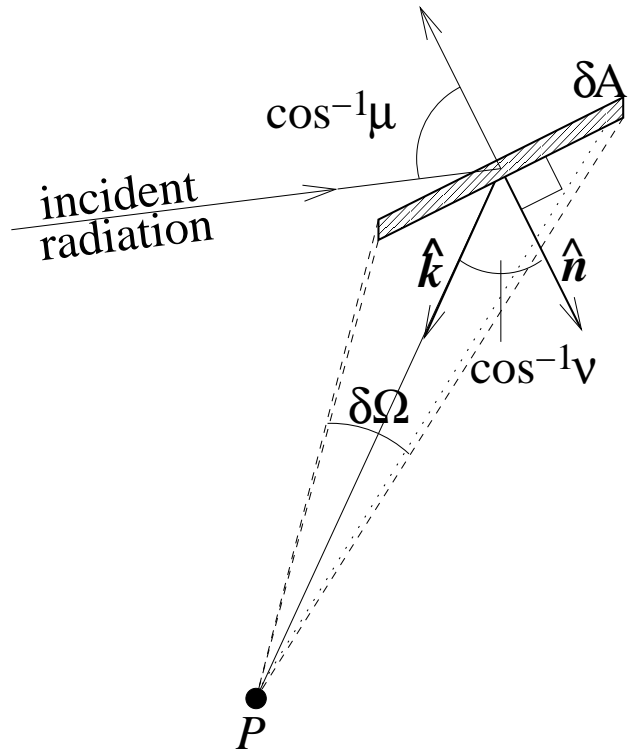}
\caption{\label{angles}Angles involved in calculation of perturbed surface.  
See text for details.}
\epsscale{1.0}
\end{figure}

Now we consider radiative transfer on a perturbed disk.  
A point $P(x,y,z)$ within 
the disk ``sees'' radiative flux coming from the surface of the disk.
Each area element $\delta A$ contributes a 
solid angle $\delta\Omega$ of flux to $P$.
\figref{angles} shows a schematic of such an area element.  
The angle of incidence of radiation on this surface element is 
$\cos^{-1}\mu$.  If $\mathbf{k}$ is the vector from $\delta A$ to $P$ and 
$\mathbf{\hat{n}}$ is the unit normal to $\delta A$, then 
$\nu$ is the cos of the angle between $\mathbf{\hat{n}}$ and $\mathbf{k}$.

We approximate the flux contribution from $\delta A$ as if 
it were a surface emitting isotropically with intensity 
\[ I = \frac{F_d(\tau_d,\mu)}{\pi}. \]
and the corresponding contribution to the total flux is 
$ d\mathbf{F} = I \,\mathbf{\hat{k}}\; d\Omega.$
Then the total flux at $P$ over all surface 
contributions is 
\begin{equation}
\mathbf{F}_{\mathrm{tot}} = 
\frac{1}{\pi} \int F_d \,\mathbf{\hat{k}}\; \delta\Omega.
\end{equation}
Since $\nabla$ is a linear operator,
\begin{equation}
\frac{1}{\kappa_d\rho}\nabla\cdot\mathbf{F}_{\mathrm{tot}} =
	\frac{1}{\pi} \int \frac{d F_d}{d\tau_d} \,
		\mathbf{\hat{n}} \cdot \mathbf{\hat{k}}\; \delta\Omega =
	4\pi\left(
	\frac{1}{\pi} \int J_d\, \nu \; \delta\Omega -
	\frac{1}{\pi} \int B\, \nu \; \delta\Omega \right).
\end{equation}
Defining 
\begin{eqnarray}
J_{\mathrm{tot}} &=& \frac{1}{\pi} \int J_d(\tau_d,\mu)\,\nu\;\delta\Omega 
\quad \mbox{and} \\
B_{\mathrm{tot}} &=& \frac{1}{\pi} \int B(\tau_d,\mu)\,\nu\;\delta\Omega 
\label{btot}
\end{eqnarray}
we see that $\mathbf{F}_{\mathrm{tot}}$, $J_{\mathrm{tot}}$, 
and $B_{\mathrm{tot}}$ satisfy equations (\ref{d0}) and (\ref{d1}), 
and we can calculate the perturbed temperature as 
\begin{equation}
T_r = \left(\frac{\pi B_{\mathrm{tot}}}{\sigma_B}\right)^{1/4}.
\end{equation}

For a plane-parallel surface, $\tau_d$ and $\mu$ are constant, and 
$\nu = \cos \theta$,
where $\theta$ is the angle with respect to the surface normal.
We then recover the solution for the plane-parallel case.

This method of calculating radiative transfer can be applied to 
any three-dimensional disk configuration, including perturbations 
induced by protoplanets.  Sufficiently massive protoplanets will 
open a gap in the disk, in which case the effects of shadowing and 
illumination across the gap can be calculated for all azimuthal angles.  
In the following section, we will calculate radiative transfer on 
small perturbations induced by 
planets too small to open a gap, but large enough to act on the 
disk locally.

\section{Hydrostatic Equilibrium}

Now we consider a perturbation induced by the gravitational potential 
of a protoplanet on a circumstellar disk.  We shall consider only 
the gravitational effects of the protoplanet on hydrostatic 
equilibrium and assume that any resonant effects are washed out by 
gas drag.  

\subsection{Vertical Density Profile}

In order to find the vertical structure of a non-self-gravitating 
gaseous disk orbiting a central star, 
we solve the equation of hydrostatic equilibrium:
\begin{equation}
\nabla P = - \rho \nabla \Phi
\end{equation}
where $P$ is the gas pressure of the disk and $\Phi$ is the gravitational 
potential of the central star.  Since we are interested in the vertical 
structure, we consider only the $z$ components and assume that the disk is 
isothermal in the $z$ direction.  For a disk without a perturbing 
protoplanet, \( \Phi = GM_{\star}/a \) where $M_{\star}$ and $a$ 
are the mass and distance of the star, respectively.  
If $c_s$ is the isothermal sound speed, then 
\begin{equation}
c_s^2\,\frac{d\, \ln \rho}{dz} = 
- \frac{GM_{\star}}{a^3} z.
\label{thediffeq}
\end{equation}
If we assume that $a\gg z$ and define 
\( h \equiv c_s\,a^{3/2}/( GM_{\star})^{1/2}
= (c_s/v_{\phi}) a, \)
then 
\begin{equation}
\rho(z) =  \rho_0 \exp\left(-\frac{z^2}{2 h^2}\right),
\label{unperturbed}
\end{equation}
i.e.~the density has a Gaussian distribution with a scale height 
of $h$ which is determined by the local temperature. 

The insertion of a planet into a passive disk adds an additional term 
to $\Phi$, representing the potential due to the planet:
\[ \Phi = GM_{\star}/a + G m_p/r_p \]
where $m_p$ and $r_p$ are the mass and distance of the planet.
Assuming that the planet is in the midplane ($z=0$),
Eq.~(\ref{thediffeq}) becomes
\begin{equation}
\frac{d\, \ln \rho}{dz} =  -\frac{z}{h^2}
	- \left(\frac{M_p a^3}{M_{\star} h^2}\right) \frac{z}{r_p^3}.
\end{equation}
The Hill radius is defined as 
\( \rhill \equiv (M_p/3M_{\star})^{1/3} a,\)
and this equation integrates to 
\begin{equation}
\rho(z) = \rho_1 \exp\left[-\frac{z^2}{2 h^2}
	+ \frac{3\rhill^3}{h^2 \, r_p}
	\right]
\end{equation}
where \(r_p = \sqrt{x^2 + y^2 + z^2}\), 
with the planet as the coordinate origin.
So long as $\rhill \ll h$ and $|z|>\rhill$, 
the effect of the planet is a small 
perturbation on the density profile.  

We normalize this equation by matching 
the density at $z=0$ to the unperturbed density, thus
\begin{equation}
\rho_p(z) = \rho_0 \exp\left[-\frac{z^2}{2 h^2}
        + \frac{3\rhill^2}{h^2}
	  \left(\frac{\rhill}{\sqrt{x^2+y^2+z^2}}
	-\frac{\rhill}{\sqrt{x^2+y^2}}\right)\right]
\label{pertdens}
\end{equation}

\begin{figure}
\plottwo{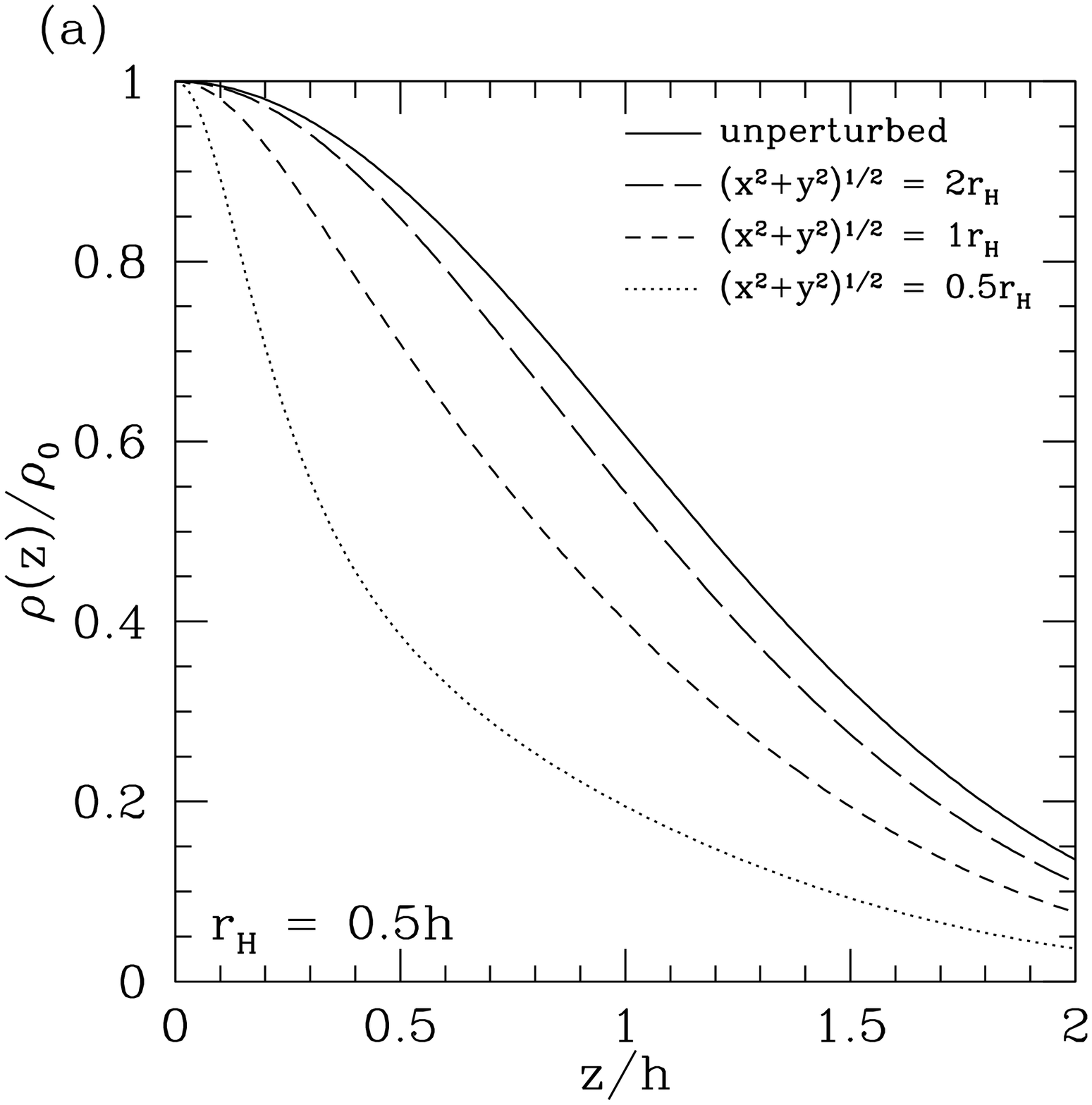}{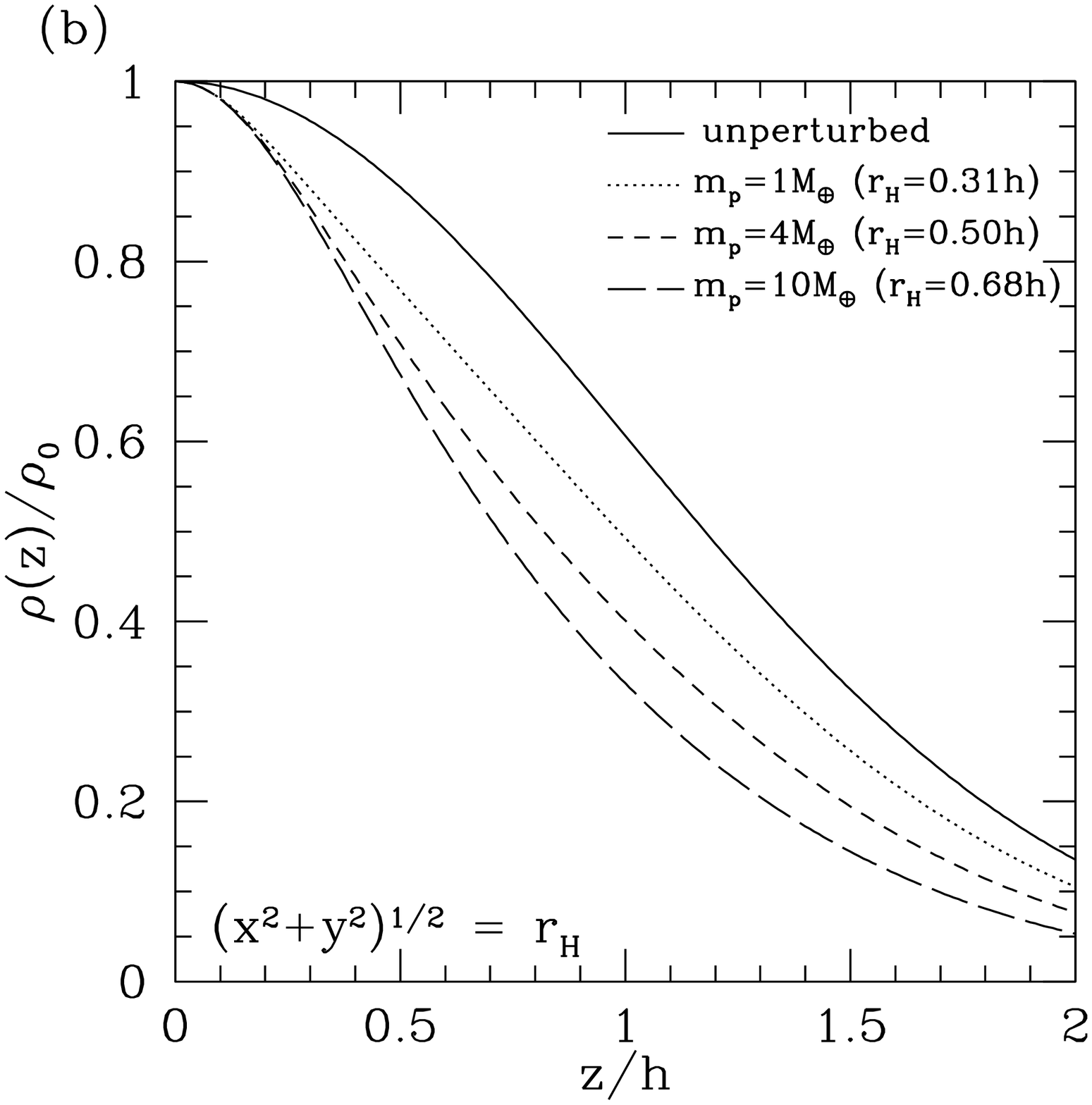}
\caption{\label{zprof}Perturbed vertical density profile for varying 
parameters.  Density is normalized to the midplane density.  Both 
plots are to the same scale and show the unperturbed density profile 
as a solid line.
(a) Density profile for fixed Hill radius and 
varying values of $(x^2+y^2)^{1/2}$ as indicated.
(b) Density profile for $(x^2+y^2)^{1/2}$ equal to the 
Hill radius and varying planetary masses.
}
\end{figure}

Note that there is a singularity ar $r_p = 0$.   \figref{zprof} shows 
the shape of this density profile. 
In \figref{zprof}a, we hold $\rhill/h$ fixed at $0.5$ and vary
$\sqrt{x^2+y^2}$.  
For a disk with $h=0.04\AU$ at $1\AU$ around a central star of
$0.5\:\msun$, this corresponds to a protoplanetary mass of
$4\:\mearth$.  For
$(x^2+y^2)\gtrsim\rhill^2$, the perturbed density profile does not
deviate significantly from the unperturbed density profile.  For
$(x^2+y^2)\lesssim\rhill^2$, the density profile develops a sharper
profile at $z<h$.  
In \figref{zprof}b, we hold $\sqrt{x^2+y^2}$ fixed at the Hill radius
and vary the mass of the protoplanet, effectively varying $\rhill$.
Larger protoplanets have the effect of decreasing the overall density 
profile, whereas position controls the shape of the profile, 
particularly close to the protoplanet.  
Since the flow of gas within the Roche lobe cannot
be adequately described hydrostatically, we exclude the region
$\sqrt{x^2+y^2} \lesssim \rhill$ from our analysis.

The normalization on \eqref{pertdens} is somewhat arbitrary, but 
at the disk heights that are of interest to us, the overall normalization 
has small effect.  We demonstrate this as follows: 
as $z\gg\rhill$, 
\eqref{pertdens} becomes
\begin{equation}
\rho_p(z\gg\rhill) \approx \rho_0
\exp\left(-\frac{z^2}{2 h^2}-\frac{3\rhill^3}{h^2\sqrt{x^2+y^2}}\right),
\end{equation}
that is, a Gaussian with respect to $z$ with some normalization 
that depends on $\rhill$ and $\sqrt{x^2+y^2}$.
The height of the photosphere, $H$, is generally several times the 
pressure scale height.  If we take $H/h=5$, then the change in 
density with vertical distance is 
\begin{equation}
\label{drho2}
\frac{\Delta\rho}{\rho} = 
\left|\frac{\Delta z}{\rho}\frac{d\rho}{dz}(z=H)\right|
= \frac{25\Delta z}H.
\end{equation}
So, a density change by a factor of $1/2$ corresponds to a shift in 
vertical height of 2\%.  In other words, the density gradient in the 
photosphere is so high that changes in the normalization of the 
density will not significantly affect our results. 
 
\subsection{Photosphere Height}
\label{photosphere}
We calculate the height of the perturbed photosphere from the 
perturbed density profile as the isodensity contour at the 
density of the unperturbed photosphere.
That is, if $H_0$ is the unperturbed photosphere 
height and $H=H_0(1-\epsilon)$ is the new photosphere height, then 
we set 
\[
\rho_d(H_0) = \rho_p(H)
\]
or
\begin{equation}
-\frac{H_0^2}{2} =
-\frac{H^2}{2} + 3\rhill^3 
	  \left(\frac{1}{\sqrt{x^2+y^2+H^2}}-\frac{1}{\sqrt{x^2+y^2}}\right).
\end{equation}
Setting \( \sqrt{x^2+y^2} = \gamma H_0 \), then
to first order in $\epsilon$

\begin{equation}
H \approx H_0 \left\{1 - 
	3\left(\frac{\rhill}{H_0}\right)^3
	\left[ \frac{1}{\gamma} - \frac{1}{\sqrt{\gamma^2+1}} \right] 
	\right\} 
\end{equation}
so that the depth of the perturbation scales as $(\rhill/H_0)^3$.  
For typical disk models, the ratio of the photosphere height to the 
thermal scale height, $H_0/h$, is between 3 and 5, depending on the 
distance from the star and is often taken to be a constant 
\citep{kenyon,CG,vertstruct}.  Restating the scaling of the 
depth of the perturbation as $\propto(\rhill/h)^3$, we see that this 
agrees with the estimate made in \citet{sas_lec}.

\begin{figure}
\plotone{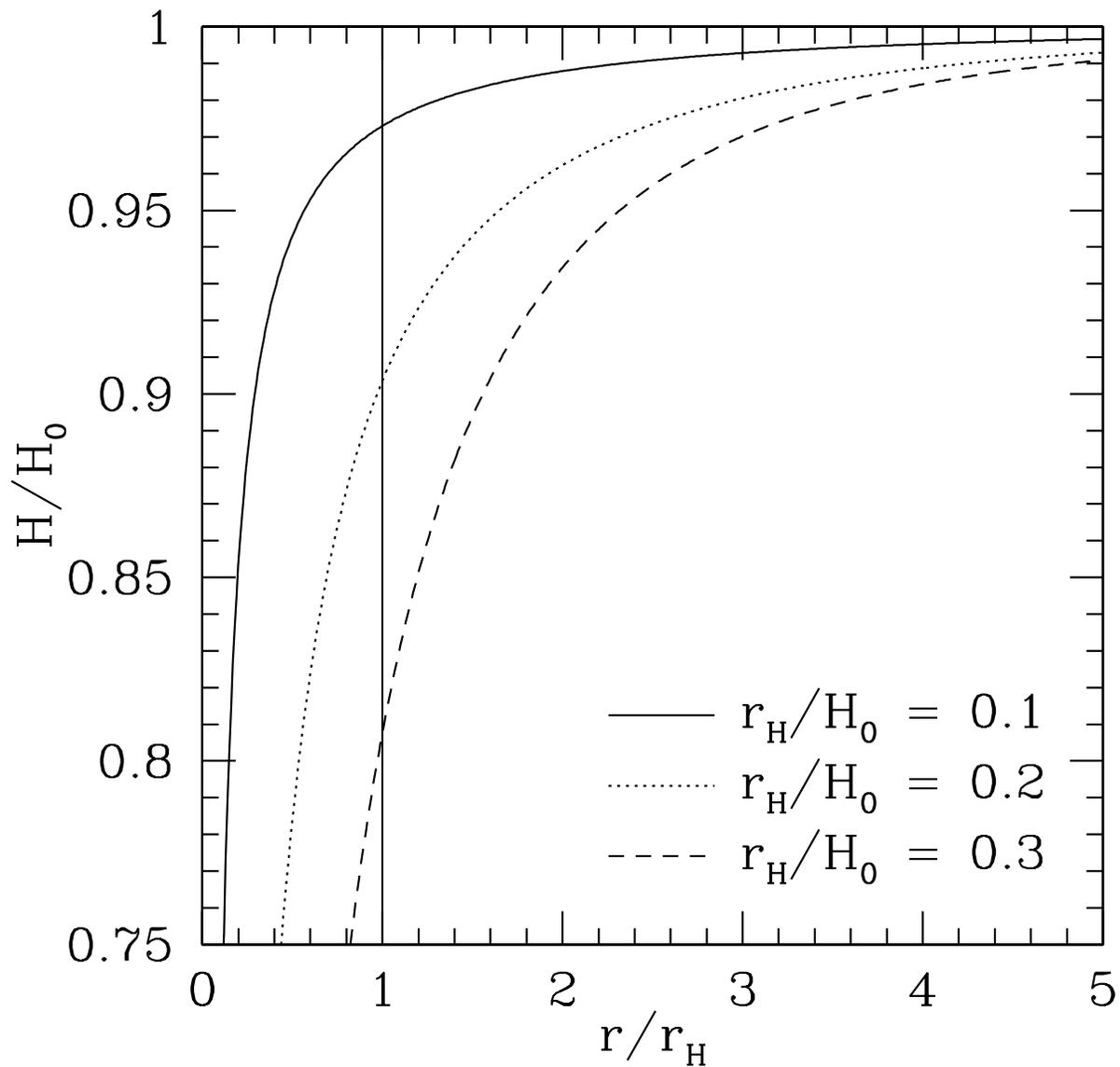}
\caption{\label{hprof}Perturbed photosphere height as a function of 
distance in the $xy$-plane, $r$, for varying Hill radii.  The solid vertical 
line indicates where $r=\rhill$.
}
\end{figure}

The shape of this perturbed photosphere is shown in \figref{hprof} 
for varying values of $\rhill$.  The perturbation is small for $r>\rhill$, 
and there is a singularity at $r=0$.  As the Hill radius increases, 
the depth of the perturbation also increases.  
Note that if $H_0/h=5$, then $\rhill/H_0=0.2$ means that the Hill 
radius is comparable to the disk pressure scale height and
the perturbation is no longer small.  Protoplanets this massive are 
likely to open a gap in the disk rather than simply inducing a 
perturbation to the photosphere.

\section{Disk Shear}

The differential rotation of the disk causes the disk material 
near the protoplanet to move through the perturbation at a rate 
that depends on the velocity with respect to the protoplanet's 
orbit.  Material moving along a given streamline in the disk 
may pass through the perturbation too quickly to experience 
significant heating and cooling, thereby diminishing the effect 
of the perturbation on the temperature structure of the disk.

The velocity of the disk material with respect to 
the protoplanet is 
$v \simeq a(\Omega-\Omega_p)$ where $\Omega$ and $\Omega_p$ are 
the orbital angular velocities of the disk material and the protoplanet,
respectively, and $a$ is the orbital radius of the protoplanet.
The orbital angular velocity of a gaseous disk is 
\begin{equation}\label{streamvel}
r^2\Omega^2 = \frac{GM_{\star}}{r} + \frac{1}{\rho}\frac{dP}{dr}.
\end{equation}
In thin disks, such as we are investigating, the pressure gradient 
term is small so the orbital velocity is nearly equal to the 
Keplerian velocity.  

The heating/cooling rate of the disk material can be expressed as 
\begin{equation}
C \frac{\partial T}{\partial t} = F - \sigma T^4
\end{equation}
where we take $F = \sigma(T_v^4+T_r^4)$ and $C$ is the specific 
heat per unit surface area of the disk.
We shall adopt a specific heat of $C = k\Sigma/m$ where $k$ is the 
Boltzmann constant, $\Sigma$ is the total surface density, and 
$m$ is the mean molecular mass.  

We shall assume that the streamlines of the disk material follow 
\eqref{streamvel}, and that they are not perturbed by the 
protoplanet's graviational potential.  This is an adequate approximation 
for disk material outside the protoplanet's Hill radius.
Along each streamline, we calculate the total radiative flux, $F$, 
at a given position, and using the velocity of the streamline 
with respect to the protoplanet along with heating/cooling 
rate, we can calculate the steady state temperature at each position 
in the disk.

\section{Model Parameters}

Dust properties can be parametrized by $\alpha$, $\kappa_s$, and $q$.
Then, the temperature profile in a plane-parallel disk is determined 
completely by $F_v$, $E_0$, $\mu_0$, and $\tau_d$, where $\tau_d$ is 
measured normal to the surface of the photosphere.  
Stellar properties determine $F_v$ and $E_0$, 
and the disk profile determines $\mu_0$.
If $H_0 \propto r^{\xi}$, then the incident angle is determined by 
\begin{equation}
\mu_0 = \frac{ (\xi-1) H_0/a }{
[1+(H_0/a)^2][1+\xi^2(H_0/a)^2]}
\end{equation}

A perturbation changes the local angle of incidence and optical depth.  
In particular, without plane parallel symmetry, there is no single 
optical depth that parametrizes the distance below the surface.  
Instead, we sum over different lines of sight, using the optical depth 
along each line of sight 
to determine the contribution to the disk flux as in \eqref{btot}.  
The local angle of incidence depends only on the geometry of the 
perturbed surface, while the optical depth also depends on the 
density structure along the line of sight.  
The shape of the perturbed surface depends on $\rhill$ and $H_0$, 
and the density structure depends on the disk pressure scale height, $h$.

\subsection{Fiducial Model}

Motivated by observations of T Tauri stars, we will assume 
for our fiducial model that the central star 
has mass $M_{\star} = 0.5\;\msun$, radius $R_{\star} = 2 R_{\odot}$, 
and effective temperature $T_{\star} = 4000$ K, 
and that the protoplanet is at $a = 1\AU$ from the star.  
We assume an accretion rate of 
$\dot{M}_a = 10^{-8}\;\msun\,\mbox{yr}^{-1}$, so that 
$T_v = 72$ K at $\tau_d = 2/3$ at 1 AU.  
With these parameters, heating in the photosphere will be dominated 
by stellar irradiation rather than viscous heating.  
Near the midplane, however, viscous heating will dominate.  
Since we only consider the photosphere in our model, 
the calculation of the midplane temperature is outside the scope 
of this paper.  

We will assume that the fraction of scattered radiation is $\alpha = 0.28$, 
with opacity in the optical of 
$\kappa_s = 400\;\mbox{cm}^2\mbox{g}^{-1}$, and 
ratio of opacities $q = 20$.

We shall assume that $H_0 = 0.16\AU (r/1\AU)^{9/7}$,
consistent with \eg \citet{dalessio2}.  
In calculating the detailed density structure in the photosphere, 
we shall assume that $h=3.4\times10^{-2}\AU$ so that $H_0/h = 5$.  
A protoplanet with mass $m_p = 1\;\mearth$ ($10\:\mearth$) will have a 
Hill radius of $1.26\times10^{-2}\AU$ ($2.72\times10^{-2}\AU$).  
As shown by \citet{bate}, planets with masses $\lesssim 30\:\mearth$ 
are not massive enough to open a gap, and our perturbative 
treatment is valid.  We have calculated models with 
$\:1\mearth$, $10\:\mearth$, and $20\:\mearth$ planets, and present the 
results below.

\subsection{Numerical details} 

We define our coordinate axes so that 
$x$ is aligned with the radial direction, 
$y$ is aligned with $\phi$ (i.e.~the direction of planet's orbit), 
and $z$ is perpendicular to the orbital plane.
The coordinate origin is set at the surface of the disk above 
the planet position, so that the planet's coordinates are 
$(0,0,-H_0)$.  

We calculate the shape of the disk surface in the $x$ and $y$ directions 
over a range \( -10\rhill \leq x< 10\rhill \) and 
\( -10\rhill \leq y< 10\rhill \), 
and use this surface to numerically 
integrate temperatures within the photosphere, to a depth of 
$z(\tau_d=2/3) = -0.011\AU$.
The temperature is calculated at each point within a 
$64\times64\times64$ grid.

When summing over surface elements to find $B_\mathrm{tot}$ as in 
\eqref{btot}, we can approximate 
$\delta\Omega\approx\nu\delta A/\ell^2$ 
for $\delta A \ll \ell^2$, where $\ell$ is the distance between 
the point $P$ and $\delta A$.  However, 
for points close to the surface, this approximation breaks down.  
In addition, $\nu$ may have large excursions over $\delta A$.  
For this reason, we assume that each $\delta A$ is approximately planar, 
so that $B(\tau_d,\mu)$ does not vary, but we calculate 
$\int_{\delta\Omega} \nu\: d\Omega$ analytically from the limits 
of $\delta A$.  In other words, we calculate
\begin{equation}
B_\mathrm{tot}
 = \frac{1}{\pi} \sum B(\tau_d,\mu) \int\limits_{\delta\Omega} \nu\: d\Omega
\end{equation}
and solve for the temperature.

\section{Results}

\begin{figure}
\plotone{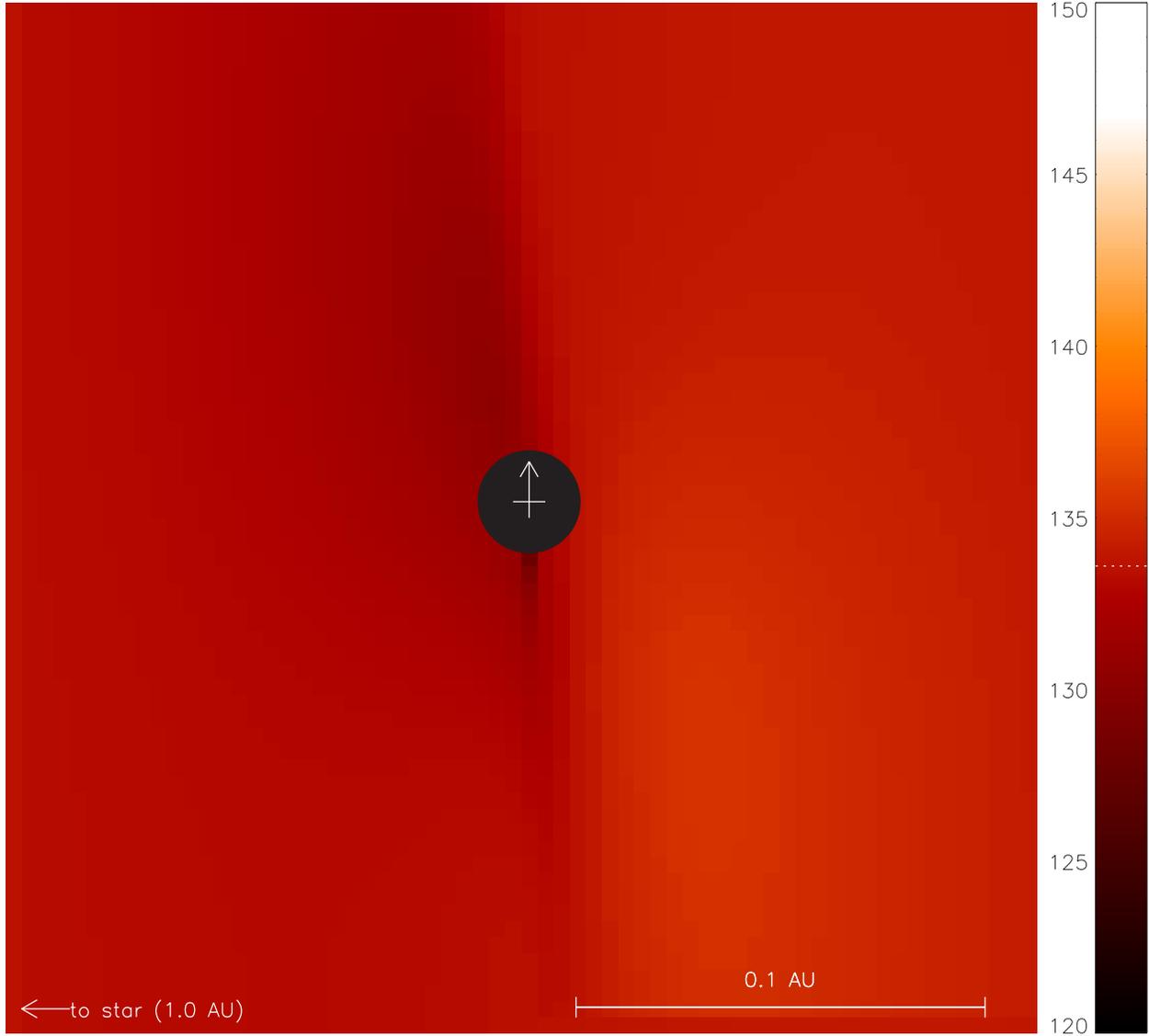}
\caption{\label{planet01}Temperature profile of the disk photosphere 
near a protoplanet of mass 1 $\mearth$ 
at $\tau_d = 2/3$ in the $xy$-plane.  
The colors indicate the temperature in Kelvin, scaled as shown in 
the colorbar. 
The cross in the center represents planet position, 
with the arrow indicating the direction of the planet's orbit.
The blacked-out circle indicates the extent of the Hill radius.  
The dotted white line 
in the color bar indicates the unperturbed temperature at 
$\tau_d=2/3$.
}
\end{figure}

\begin{figure}
\plotone{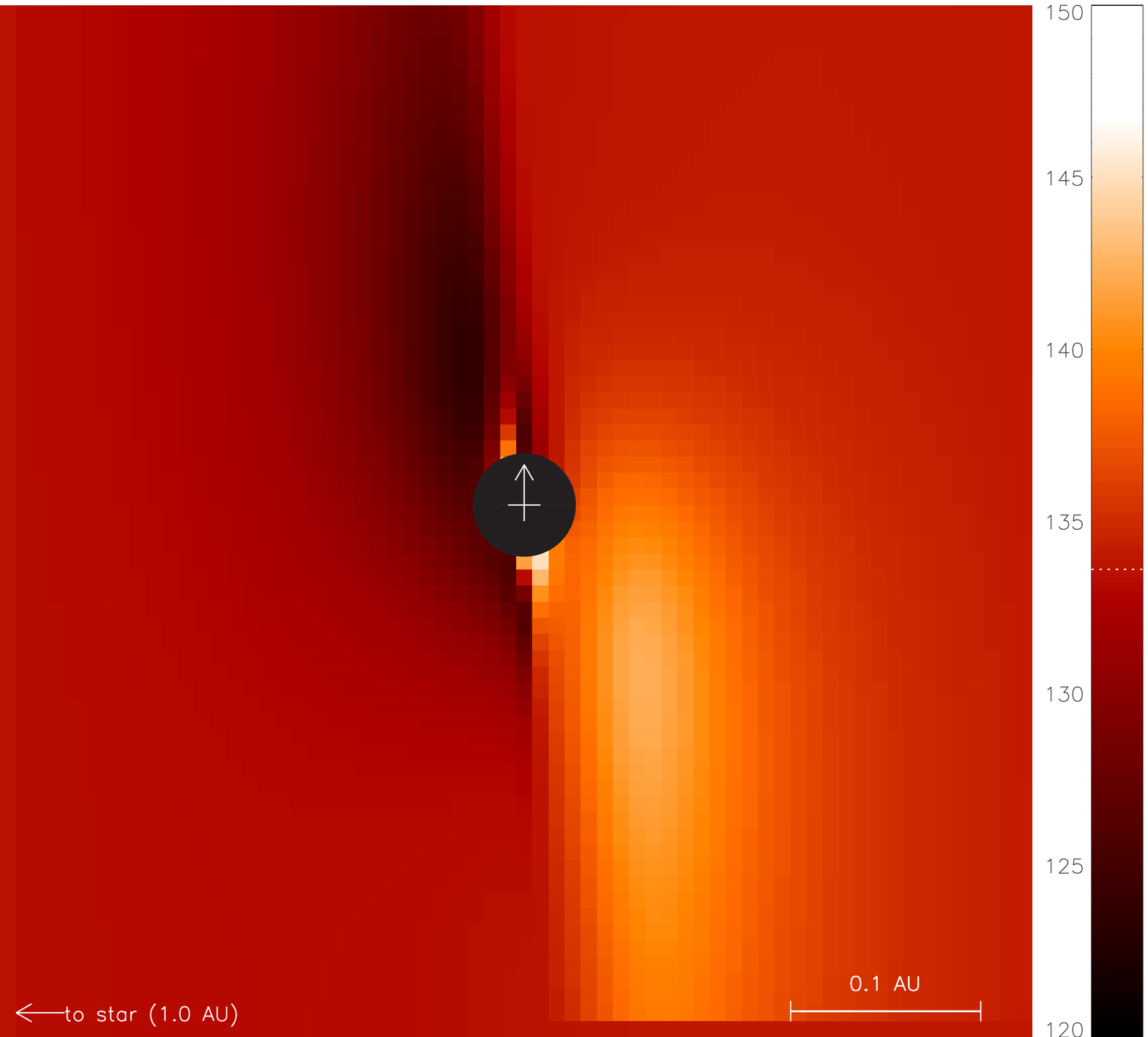}
\caption{\label{planet10}Same as \figref{planet01}, for a 
protoplanet of mass 10 $\mearth$.  Note that the dimensions 
on the figures scale with the Hill radius, so the physical sizes of 
the plots are $10^{1/3}$ larger than in \figref{planet01}.
}
\end{figure}

\begin{figure}
\plotone{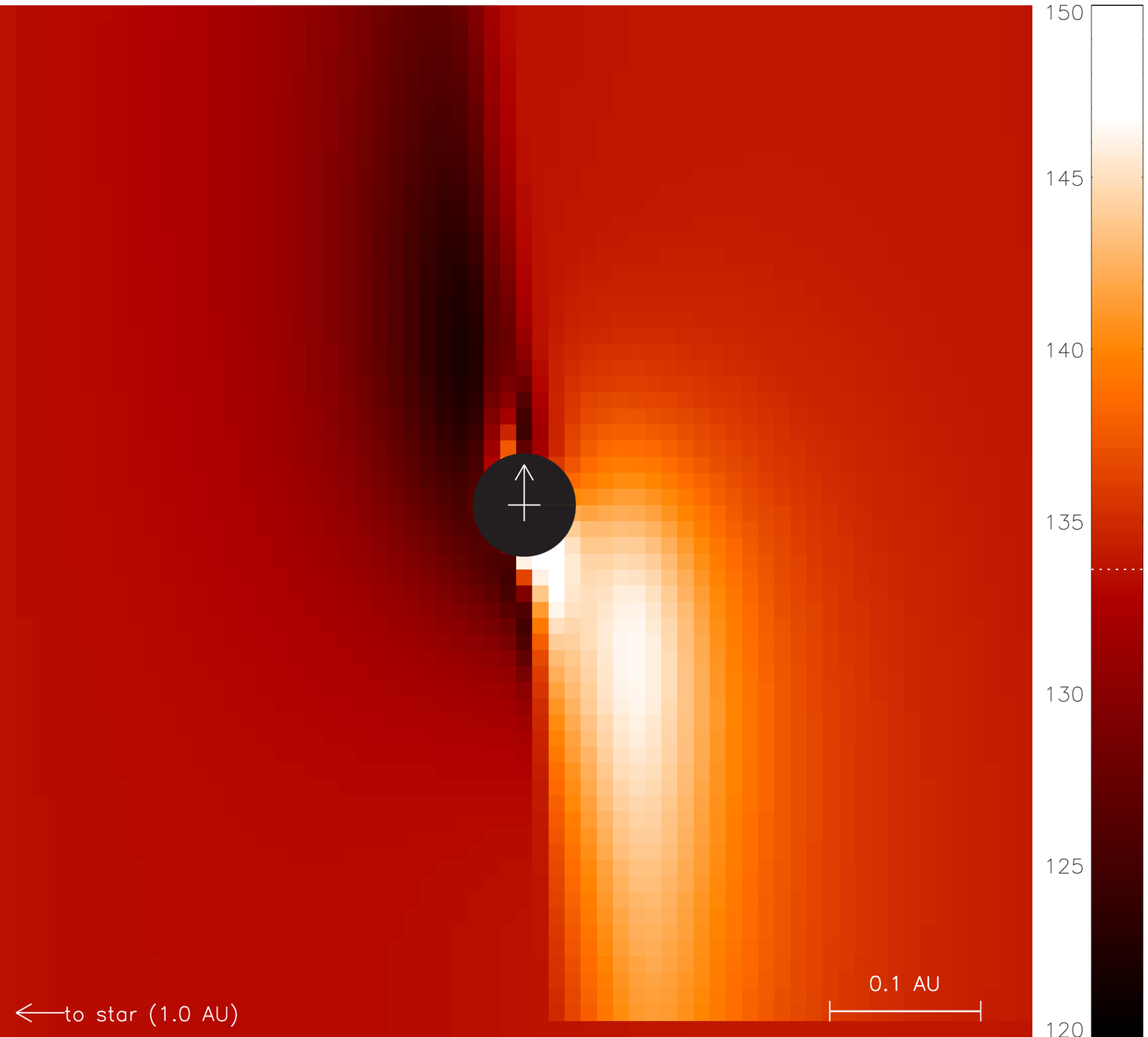}
\caption{\label{planet20}Same as \figref{planet01}, for a 
protoplanet of mass 20 $\mearth$.}
\end{figure}

Figures \ref{planet01}, \ref{planet10},  and \ref{planet20} 
show the spatial distribution of temperatures 
in the photosphere of the fiducial model for $1\:\mearth$ 
$10\:\mearth$, and $20\:\mearth$ planets,
respectively.  The colors represent temperatures in Kelvin, 
as indicated by the colorbars.
The horizontal axis indicates increasing radial distance from the star 
so that the photosphere is illuminated from the left.
The cross-section is taken at the $\tau_d=2/3$ surface. 
The bottommost row and leftmost column show the
unperturbed temperature of $\tunpert$ for comparison.  This temperature
is also indicated in the colorbar by a dotted white line.  

The direction of the planet's orbit is upward on the plots, so 
the interior disk material moves faster, and the 
exterior disk material moves slower than the planet.  
This causes the heated and cooled regions to shear out.  The 
positions of minimum and maximum temperatures are offset from 
the planet's position since it takes some time for the 
heated/cooled material to come back to the equilibrium 
temperature. 

For a $1\:\mearth$ protoplanet, 
the shadowing and illumination at the surface results in a slight
temperature change, with the area to the right of the
planet reaching a maximum temperature of $\tsmmax$, and the area to 
the left cooled to a minimun of $\tsmmin$ 
-- a change of $-\losm\%$ to $+\hism\%$.  
Thus, we see that an Earth-mass planet is too small
significantly affect the interior temperature of the disk.

A larger planet will create a larger perturbation, which will lead to
a greater temperature change.  
\figref{planet10} and \figref{planet20} show the temperature distribution
in the photosphere for a $10\:\mearth$ planet and $20\:\mearth$ planet,
respectively.  
The plots are scaled to the Hill radius, which goes as 
$m_p^{1/3}$, so the physical range in Figures \ref{planet10} and 
\ref{planet20} is actually larger than that in \figref{planet01}.  
The high temperatures close to the Hill radius are due to the low
density in this region, which gives a low optical depth to the
illuminated surface.  However, since this region is not in hydrostatic
equilibrium, we exclude it from our analysis.  Outside the Hill 
radius, the temperatures at $\tau_d=2/3$ for a $10\:\mearth$ protoplanet 
range from $\tlgmin$ to $\tlgmax$, or by $-\lolg\%$ to $+\hilg\%$.  
For a $20\:\mearth$ protoplanet, which is close to but not at the limit 
where a gap in the disk can form, the temperature range increase 
slightly, with a minimum of $\txlmin$ and maximum of $\txlmax$, 
or by $-\loxl\%$ to $+\hixl\%$.  
Thus, a protoplanet with mass 10 $\mearth$ 
or larger may have a significant effect on the local
temperature structure of a protoplanetary disk.

\section{Discussion}

As planet-forming disks are likely to be passive \citep{calvet2002}, 
we have explored
stellar irradiation effects due to a disk surface perturbation
induced by an embedded protoplanet. We find that even a relatively small
protoplanet (in the range from 1 to 20 Earth mass) produces enough
of a perturbation to cause shadowing. 
In particular, we find that in a standard passive disk accreting 
at $10^{-8}\:\msun\mathrm{yr}^{-1}$, 
the temperature variation induced by a $1\:\mearth$ planet at 1 AU 
is $\tsmd$, or only about $\dsm\%$.  However, we find that even though 
a protoplanet may not be massive enough to open a gap in a disk, it may 
still be able to significantly change the temperature structure in 
a the disk material near it.  
A $10\:\mearth$ ($20\:\mearth$) planet can induce a 
temperature variation of $\tlgd$ ($\txld$), or $\dlg\%$ ($\dxl\%$).

This results in a local temperature
perturbation which is large enough to affect the local properties of
the disk and the accretion rate, and possibly the migration rate
of the protoplanet.  
Although the perturbation in temperature may be 
too subtle to directly observe, it can have serious consequences for 
planet growth and migration.

\citet{wardA} has shown that 
the rate of Type I migration is strongly dependent upon the temperature 
gradient of the disk.  If the temperature decreases with distance from 
the star, then the total net torques result in inward migration 
of the secondary.  However, if $k = -d\ln T/d\ln r \lesssim -1$, then 
the opposite is true and the secondary will migrate outward.  
The temperature perturbation near a protoplanet locally 
decreases the value of $k$, and may result in halting or even 
reversing Type I migration.  This may resolve the conundrum that 
Type I migration timescales are typically much less than observed 
disk lifetimes of $\sim10^7$ years \citep{hartmann}.
Yet to date, no simulation of planet-disk interactions accounts 
for local temperature pertubations as described here.  

We have not yet considered 
the behavior of the dust (as a function of size) in
the perturbed region.  We expect the dust to remain coupled
to the gas and compress its scale height accordingly. In
fact crude estimates of dust sedimentation in the vicinity
of the planet confirm that. However, a detailed estimate of
the sedimentation timescale is necessary in order to evaluate
the amount of residual dust, \eg swept by stellar radiation
from the inner edge, and its total opacity. These are issues 
we plan to address in future work.  

In a forthcoming paper, we will undertake a parameter study 
of our model for radiative transfer on a perturbed disk.  
By varying attributes of the disk, star, and protoplanet, we 
will investigate the possibility that a relatively small protoplanet 
which is not large enough to open a gap in a disk might still be 
able to affect disk structure.

\acknowledgements
We thank Nuria Calvet for numerous helpful comments in the preparation 
of this paper.  

\bibliographystyle{apj}
\bibliography{apj-jour,/home/hjang/Thesis/Papers/planets}

\end{document}